# Android Based Emergency Alert Button

Dhrubajyoti Gogoi, Rupam Kumar Sharma

*Abstract*— *Android is a java based operating system which runs on the Linux 2.6 kernel. It's lightweight and full featured. Android applications are developed using Java and can be ported to new platform easily thereby fostering huge number of useful mobile applications. This paper describes about a SOS application being developed and its successful implementation with tested results. The application has target users those sections of the people who surprisingly falls into a situation where instant communication of their whereabouts becomes indispensable to be informed to certain authorized persons at remote end.*

*Index Terms*— *Gprs, SOS, security, android*

## I. INTRODUCTION

The security of women at night and at times even in day when travelling alone is a concern. On 16th December,2012 New Delhi, capital of India witnessed a heinous crime. A female physiotherapy intern was beaten and gang raped by six persons. The ambulance and other service had reached the spot late hindering emergency medical treatment. It has been observed that at times the instant communication of message of one's whereabouts precisely is a problem. This paper describes about an SOS application developed in android platform. The uniqueness of this application apart from other SOS application available is that the user need not spent time navigating inside the phone menu; unlock the screen, to trigger the service. He instead, can directly press the power button and thereby, popping up a SOS screen and user can directly click the SOS button triggering the application in the background , sending the location (latitude and longitude) to all the pre-registered phone numbers in the application. Many applications available in the market sends a custom message to the number registered but not the location of the user. In the proposed and tested application the longitude, latitude information and the general idea of the place (BTS location area) of the current position of the mobile user is appended with the custom message that had been initially set in the application and is transmitted to the phone numbers registered. This feature of the application not only helps in finding the exact location of the person in problem but also will help the police to trace the location of incident at latter time easily.

## II. EXISTING ANDROID SOS APPLICATION

There are lot many android applications available in the web today. Some are free and many need to be procured. Some of the SOS based Android Applications are listed below

### A. *SOS Emergency Support prepared by American Red Cross*

This application provides step-by-step instructions on dealing with a variety of emergencies, including choking, broken bones, strokes, allergic reactions and many more. It is a free application. It provides dozens of videos to coach a person through emergency protocols. Easy access to 9-1-1.If a person is not from the US, the application will determine what country the person is in and dial the appropriate number.[6]

### B. *Olalashe Emergency Alert Button (SOS)*

Olalashe Emergency SOS is an emergency SOS application. It allows entering in-case-of-emergency-contact from phonebook. Send SMS to contact registered that the user is in trouble. Click the widget button to trigger the application [7].

## III. PROPOSED MODEL

The proposed model is designed and implemented with the objective that it has to be user friendly and triggering of the application should take least time. The location of the user in problem should also be precisely known to all those persons whom message has been sent. The proposed model is shown in the figure 1 below. The SOS button is displayed in the home screen of the mobile to avoid waste of time navigation to the application stored somewhere else. Pressing the SOS button triggers the application in the background and immediately the location of the user in terms of latitude, longitude and general information of the place the user is currently in is send automatically to the registered emergency phone numbers in the application. The application for full functioning demands GPS service to be available in the handset. If the handset don't have GPS service, attempt to trigger this application will show an error message, but still sending an sms to the registered phone numbers. This feature is very useful taking those users who don't have GPS enabled handset. If the user is not triggering the sos button then the default home screen of the mobile continuous to be displayed.

## IV. PACKAGES USED

Some of the packages used to accomplish retrieving the location using GPS services are android.location.Location, android. Location. Location Listener, android. Location. Location Manager etc. The package used for sending SMS to the emergency numbers is android.telephony.SmsManager. The custom class AppPreferences.java imports preference. Preference Activity to save the numbers and addPreferencesFromResource is used for calling the emergency numbers and retrieving them from the stored directory. meant

# Android Based Emergency Alert Button

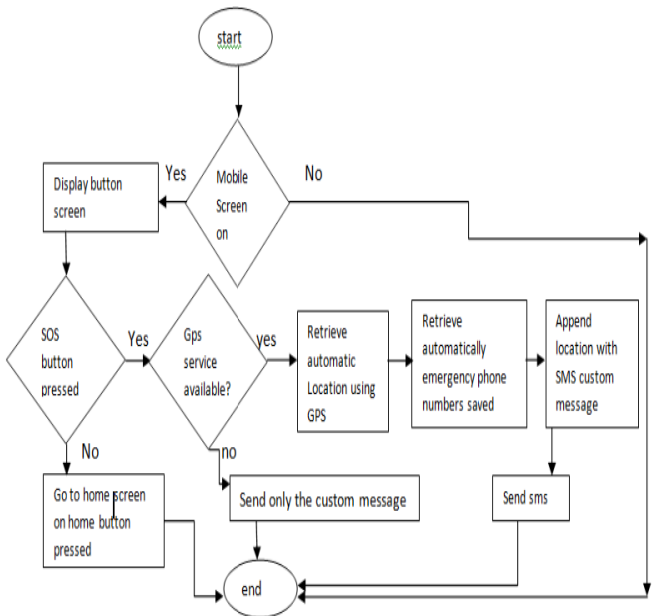

Figure 1

Another custom class BroadcastSetter.java imports the android.content.BroadcastReceiver. The file displays the SOS screen above the mobile home screen.

## V. TESTING AND OUTPUT RESULTS

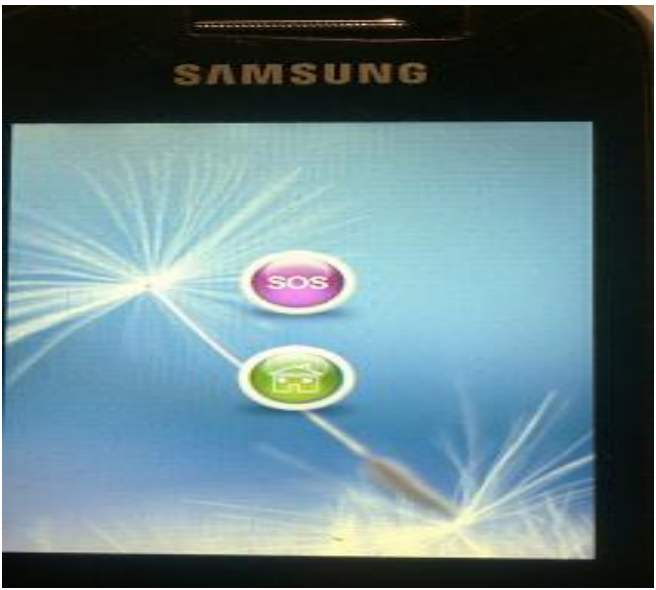

Figure: Home screen showing the SOS button

### A. Home screen showing the SOS Button.
The above figure displays the Home Screen with SOS Button. The SOS application can be immediately triggered by clicking in the SOS button, else clicking otherwise, in the home screen takes the user to the General Menu of the mobile.

### B. Application triggered Output.
16000 A/m or 0.016 A/m. Figure labels should be legible, approximately 8 to 12 point type. Sentence punctuation follows the brackets [2]. Multiple references [2], [3] are each numbered with separate brackets. The figure shows the output message received by the mobile phone. It shows the user is in Longitude:91.6 and Latitude:26.1. It also indicates that the geographic location is National Highway 37, Borjhar, Guwahati, Assam, India.

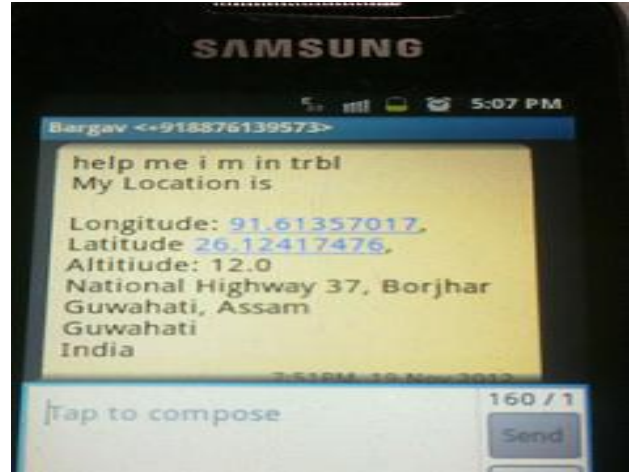

Figure: Application triggered output

## VI. CONCLUSION

This application as stated earlier can be of immense help for all those people using this application. The user neither takes time to trigger the application nor the application uses longer time to process. The application is looked forward to be incorporated with automatic location of the user using Google Map.

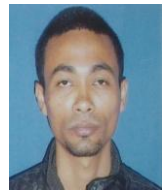

**Dhrubajyoti Gogoi** is currently pursuing his MCA degree in Don Bosco College of Engineering and Technology. He has developed a number of android applications. His area of interest is Android and Computer Network.

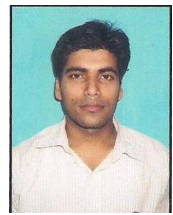

**Rupam Kumar Sharma,** Rupam Kumar Sharma is currently working as an Assistant Professor in the department of Computer Science and Information Technology under Don Bosco College of Engineering and Technology. He presently has 2 and half years of teaching experience. He has also done various training courses such as CCNA,OCA ,Semantic Training on Data Security, Data Backup, Restore and Availability. He has done his MCA under NEHU .He has one international publication in IJCSI. His current research interest is computer networks and network security.